\documentclass[useAMS,usenatbib]{mn2e}

\usepackage{graphicx}

\title[SXP 323]{SXP 323 - an unusual X-ray binary system in the Small
Magellanic Cloud \thanks{Based on observations collected at the European Southern Observatory, Paranal, Chile (ESO 71.D-0093)}}

\author[M.J. Coe et al.]{M. J.~Coe$^{1}$, I. Negueruela$^{2}$ \& V.A. McBride$^{1}$  \\
$^{1}$School of Physics and Astronomy, Southampton University, SO17 
1BJ, UK\\
$^{2}$ Departamento de Fisica, Ingeniera de Sistemas y Teora de la Seal, 
Escuela Politcnica Superior, \\
University of Alicante, Ap.99, 03080 Alicante, Spain.}
\begin{document}

\date{June 2005}

\pagerange{\pageref{firstpage}--\pageref{lastpage}} \pubyear{2002}

\maketitle

\label{firstpage}

\begin{abstract}

Spectroscopic observations taken with the VLT/UVES
telescope/instrument are presented of the unusual Small Magellanic
Cloud (SMC) X-ray binary system SXP 323 = AX J0051-733. This system
shows a clear modulation at 0.71d in long term optical photometry
which has been proposed as the binary period of this system. The high
resolution optical spectra, taken at a range of phases during the
0.71d cycle, rule out this possibility. Instead it is suggested that
this long-term effect is due to Non Radial Pulsations (NRP) in the Be
star companion to SXP 323. In addition, the spectra show clear
evidence for major changes in the (V/R) ratio of the double peaks of
the Balmer lines indicative of asymmetries in the circumstellar
disk. The complex structure of the interstellar lines are also
discussed in the context of the SMC structure.

\end{abstract}


\section{Introduction and background}

The Be/X-ray systems represent the largest sub-class of massive X-ray
binaries.  A survey of the literature reveals that of the 115
identified massive X-ray binary pulsar systems (identified here means
exhibiting a coherent X-ray pulse period), most of the systems fall
within this Be counterpart class of binary.  The orbit of the Be star
and the compact object, presumably a neutron star, is generally wide
and eccentric.  X-ray outbursts are normally associated with the
passage of the neutron star close to the circumstellar disk (Okazaki
\& Negueruela, 2001). 

X-ray satellite observations have revealed that the Small Magellanic
Cloud (SMC) contains an unexpectedly large number of High Mass X-ray
Binaries (HMXB). At the time of writing, 47 known or probable sources
of this type have been identified in the SMC and they continue to be
discovered at a rate of about 2-3 per year, although only a small
fraction of these are active at any one time because of their
transient nature.  Unusually (compared to the Milky Way and the LMC)
all the X-ray binaries so far discovered in the SMC are HMXBs, and
equally strangely, only one of the objects is a supergiant system, all
the rest are Be/X-ray binaries. A recent review of these systems may
be found in Coe et al. (2005).

The source that is the subject of this paper, SXP 323 = AX J0051-733,
is one of these SMC systems. It was reported as a 323s pulsar by
Yokogawa \& Koyama (1998) and Imanishi et al (1999). Subsequently Cook
(1998), using MACHO optical photometry, identified a 0.7d optically variable object within the ASCA
X-ray error circle. The system was discussed in the context of it
being a HMXB by Coe \& Orosz (2000) who presented some early
OGLE data on the object identified by Cook (1998) and modelled the
system parameters. Coe \& Orosz identified several problems with
understanding this system, primarily that if it was a binary then its
true period would be 1.4d and it would be an extremely compact
system. In addition, the combination of the pulse period and such a
binary period violates the Corbet relationship for such systems
(Corbet, 1986). A discussion of the object as a possible triple system
was presented in Coe et al. (2002).

In order to try and understand this unusual system better, 
high resolution optical spectra were obtained at a variety of phases
throughout the 1.4d period.

\section{Observational details}

Spectroscopy of the optical counterpart to SXP 323was obtained with
the Ultraviolet and Visual Echelle Spectrograph (UVES) mounted on the
8.2-m VLT/UT2 (Kueyen). UVES was used in the standard Diochroic \#1
mode.  The blue arm was equipped with a $2048\times4096$ 15-µm
pixel thinned EEV, providing coverage of the $\lambda\lambda$
3821--4520~\AA\ range. The red arm was equipped with a mosaic formed
by a similar EEV CCD (covering $\lambda\lambda$ 4727--5804~\AA) and a
MIT/LL CCID-20 chip (covering $\lambda\lambda$ 5818--6835~\AA). The
spectral resolution of this configuration is 40,000.

Nine 2000-s exposures, each with a SNR of $\sim$35/pixel, 
were obtained in service mode at random
times. The 9 sets are spread over the possible 1.4d period and give
a good coverage of different phases. The dates and phases of the
observations are presented in Table~\ref{tab:obs}.

\section{Spectral class}

In spite of the higher resolution and S/N ratio, the data cannot
provide a better classification than the spectrum presented in Coe et
al. (2002) as the range $\lambda\lambda$ 4500$-$4700\AA, rich in metallic
lines is absent. A co-added blue spectrum, with rather high S/N, shows
no evidence of SiIV~4089\AA or SiII~4128\AA. Though
the low metallicity of the SMC may result in weaker lines, this is
only compatible with a main-sequence classification. The spectral type
is constrained to lie in the B0.5\,V--B1\,V range. In view of the
complete absence of SiIV, we slightly favour the later
classification over the B0.5\,V preferred in Coe et al. (2002).

\begin{table}
\begin{center}

\begin{tabular}{cccc}
\hline
Obs. date & Obs. time (UT)& TJD& Phase \\
\hline
9 June 2003  & 09:12 & 2799.88 & 0.00 \\
12 June 2003 & 08:48 & 2802.87 & 0.22 \\
13 June 2003 & 09:35 & 2803.90 & 0.68 \\
23 June 2003 & 08:13 & 2813.84 & 0.71 \\
4 July 2003  & 06:21 & 2824.76 & 0.14 \\
6 July 2003  & 05:43 & 2826.74 & 0.93 \\
11 July 2003 & 05:12 & 2831.72 & 0.97 \\
28 July 2003 & 08:45 & 2848.86 & 0.17 \\
31 July 2003 & 03:56 & 2851.66 & 0.13 \\
\hline
\end{tabular}

\caption{Table of VLT observations. The phase given in the last column
is that with respect to the precise OGLE period of 0.7081d.}

\label{tab:obs}
\end{center}
\end{table}

\section{Search for binary motion}

\begin{figure}\begin{center}
\includegraphics[width=55mm,angle=-90]{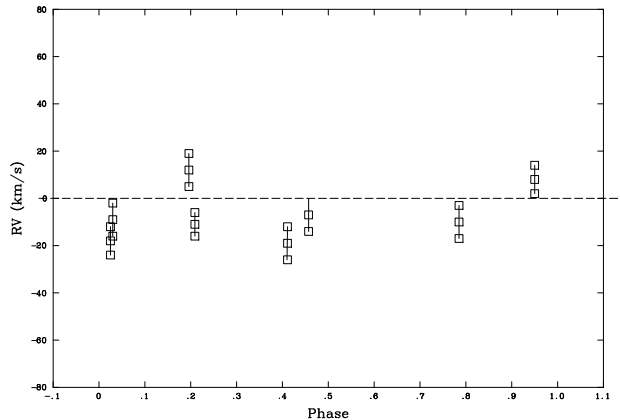}
\caption{Limits on any radial velocity motion as a function of the
0.7d phase.}
\label{fig:rv2}
\end{center}
\end{figure}

One of the main goals of the VLT observations was to search for
possible binary motion at either 0.71d or twice that value.  As can be
seen from Table~\ref{tab:obs}, a wide range of phases were covered in
the possible 0.7d or 1.4d cycle. To investigate possible radial
velocity motion in any of the line features, the spectra were
subjected to the autocorrelation analysis function RVSAO within
IRAF. The first spectra was assumed to be at a phase 0.0 and
subsequent spectra were then cross-correlated against this one. This
exercise was carried out for both the blue data (3750 - 4500\AA) and
the middle spectral range (4800 - 5700\AA). The results from the
latter are shown in Figure~\ref{fig:rv2} which set an upper limit of
$\sim$30km/s for any radial velocities. The corresponding blue band
upper limit was slightly larger at $\sim$50km/s. There was no matching
pattern between the two sets of data implying a null result for any
binary motion.

\section{Balmer line profiles}

\begin{figure}\begin{center}
\includegraphics[width=65mm,angle=-90]{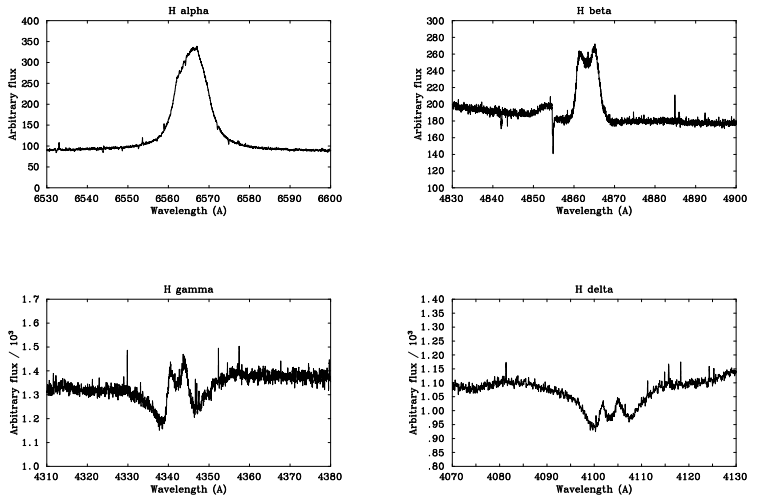}
\caption{Average Balmer line profiles for the first four of the
series. The discontinuity to the left of H$\beta$ is due to a slight
mismatch in adjacent echelle orders.}
\label{fig:balmer}
\end{center}
\end{figure}

The VLT spectra cover all the Balmer lines from H$\alpha$ to the limit
of the series. The first four Balmer lines are presented in
Figure~\ref{fig:balmer}. From this figure it is possible to see at
least two things: there is a clear progression in the strength of the
emission as we move down the series. While H$\alpha$ and H$\beta$ are
fully in emission, the emission components in upper Balmer lines are
weaker. This is perfectly normal in a Be star.
Secondly, the emission component is
clearly double peaked with V and R components.

\begin{figure}\begin{center}
\includegraphics[width=65mm,angle=-90]{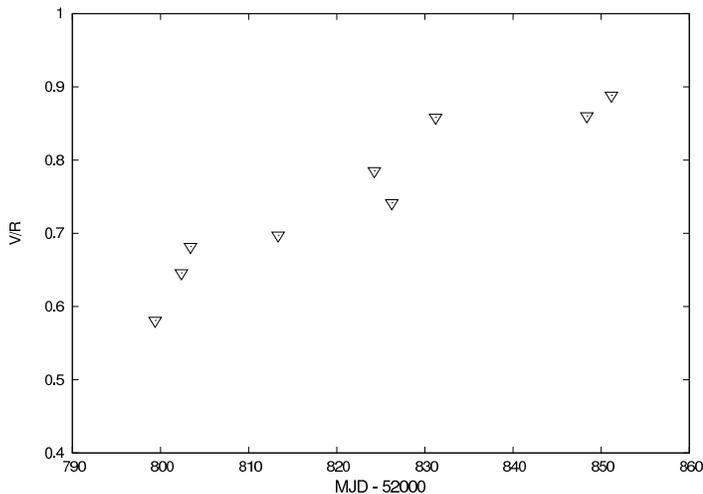}
\caption{Variations in the V/R ratio as a function of time for the H
beta line.}
\label{fig:rvhb}
\end{center}
\end{figure}

It is of great interest to use the emission features to investigate
possible V/R variations that might be indicative of global one-armed
oscillations, and hence asymmetries in the circumstellar
disk. To this end, the V/R ratio for the H$\beta$ line was measured
throughout the series of observations and the results are presented in
Figure~\ref{fig:rvhb}. It is clear from this figure that a clear
general trend exists in the ratio, the values changing from $\sim$0.6
to $\sim$0.9 over 52 days. The other Balmer lines showed a similar
trend.

\section{Interstellar line features}

\begin{table}
\begin{center}

\begin{tabular}{cccc}

Line&Feature 1&Feature 2&Feature 3\\
&km/s&km/s&km/s \\
&&& \\
CaII H & 0 & 128 & 204 \\
CaII K & 0 & 130 & 198 \\
Na D   & 5 & 132 & 200 \\
Na D   &10 & 137 & 200 \\
\hline
Average&4&132&201 \\

\end{tabular}

\caption{Table of measured interstellar line features. The error on
each measurement is approximately $\pm5 km/s$. }

\label{tab:is}
\end{center}
\end{table}

\begin{figure}\begin{center}
\includegraphics[width=55mm,angle=-90]{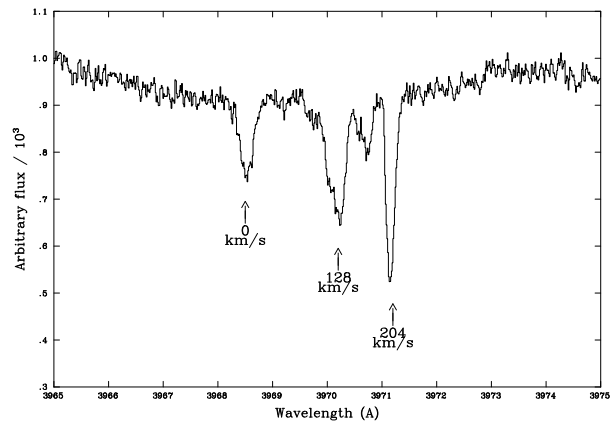}
\caption{Interstellar line features associated with the CaII H line}
\label{fig:ca1}
\end{center}
\end{figure}

The high quality, high resolution spectra permit a detailed
investigation of the interstellar line profiles in the direction of
SXP 323. Specifically two sets of IS lines were investigated: the CaII
H \& K lines (3968.5\AA ~and 3933.7\AA ~respectively), and the Na
doublet (5890.0\AA ~and 5895.9\AA).  Figure~\ref{fig:ca1} presents the
region around the CaII H line and shows the complex structure often
associated with IS lines towards the SMC (see, for example, the
extensive set of examples in Wayte, 1990). The profiles of the CaII H
\& K are almost identical to each other, and confirm the presence of
at least three significant absorption features. These are indicated in
Figure~\ref{fig:ca1} and listed in Table~\ref{tab:is}.

\begin{figure}\begin{center}
\includegraphics[width=55mm,angle=-90]{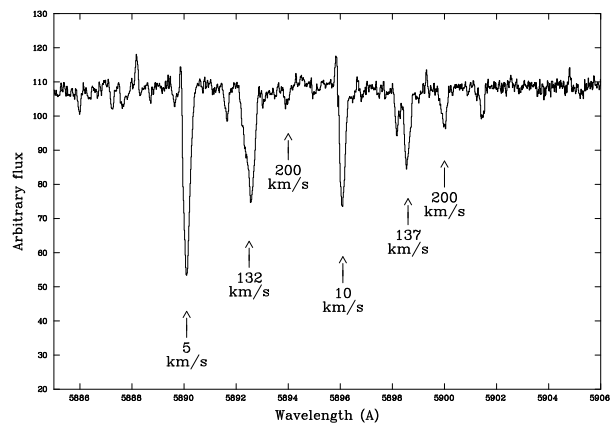}
\caption{Interstellar line features associated with the Na D lines.}
\label{fig:na}
\end{center}
\end{figure}

The region around the Na doublet is presented in Figure~\ref{fig:na}
and the values also given in Table~\ref{tab:is}.
From the Na group one can immediately see similarities and differences with
the CaII line profiles. The main similarity is the presence of at
least three major absorption features at essentially the same
velocities to those in the CaII lines. The other two features seen in
the CaII lines are weaker and cannot be ruled out as also being
present in the Na D lines.

The main difference, is the striking reversal in the relative
strengths of the the line sets between the CaII and Na features. In
the CaII lines the strongest feature is the one at the highest
velocity (204km/s) while the weakest is the one at the lowest velocity
(0km/s). The complete opposite is true for the Na lines.

\section{ Possible NRPs}

If the possibility of binary motion is excluded as an explanation for
the 0.71d modulation, then other explanations must be found. One
possible alternative would be Non-Radial Pulsations (NRP) from the Be
star. Such effects are well documented in B-type stars - see, for
example, Uytterhoeven et al, 2004. In addition, Fabrycky (2005) has
suggested a NRP explanation for two periods of 0.28d and 0.65d seen in
the OGLE data for SXP 702 (= XMMU J005517.9-723853).

\begin{figure}\begin{center}
\includegraphics[width=80mm,angle=-0]{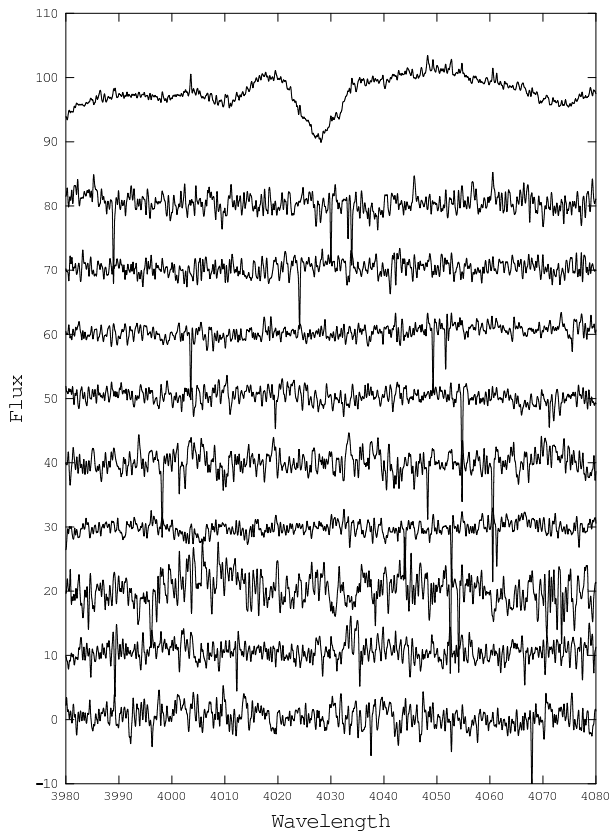}
\caption{The top line shows the average profile of the 4026\AA ~helium
line. The other lines are the individual spectra after the average
profile has been subtracted. They are arranged in order of their
occurence in the 0.71d
phase, starting from the bottom at phase 0.0. Also visible is the
4009\AA ~helium line.}
\label{fig:4026}
\end{center}
\end{figure}

The presence of strong emission components in most of the lines
renders the search for profile variability extremely difficult. The
He\,{\sc i}~4026\AA\ was selected as showing particularly little
emission contamination and its profile was studied closely to search
for any evidence of features moving in phase with the 0.71d
period. The result of this is shown in Figure~\ref{fig:4026} in which
the top line is the average profile for the He4026\AA ~line. This
profile was then subtracted from each of the individual nine spectra
and the spectra were then stacked in order of their occurence in the
0.71d phase - see Table~\ref{tab:obs}. It is not apparent from
Figure~\ref{fig:4026} that there are features left that move any
semblance of phase coherence through the spectra. Though the data are
of a sufficiently high resolution, the signal to noise limits the
detection of any such features. However, a moving feature such as that
shown in Figure 1 of Uytterhoeven et al (2004) should have been
apparent.

\begin{figure}\begin{center}
\includegraphics[width=80mm,angle=-0]{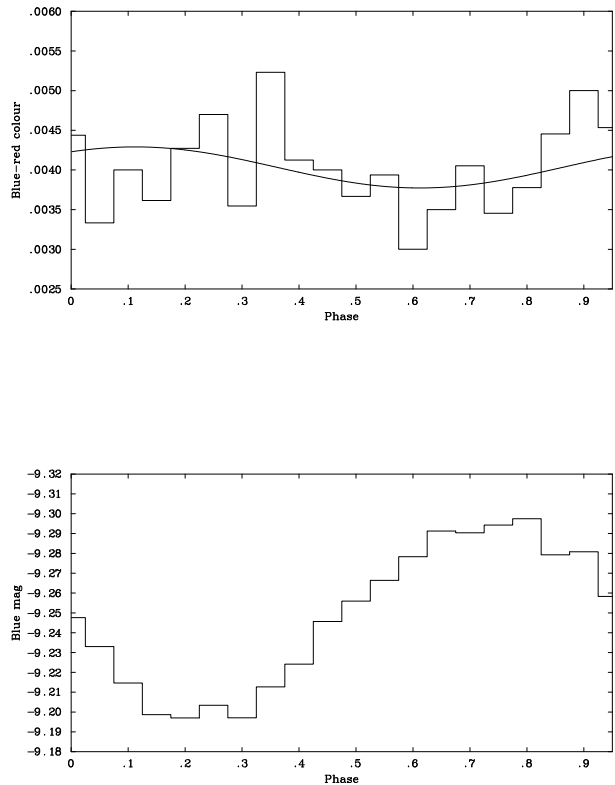}
\caption{Top panel : MACHO colour versus the phase of the 0.71d
modulation; the line through the histogram is that of the best-fit
sine wave. Lower panel : the MACHO blue data folded at the same 0.71d
period and with the same phase as the top panel.}
\label{fig:col}
\end{center}
\end{figure}

One other signature of NRP behaviour would be a temperature effect
during the pulse cycle. To investigate this, the MACHO data on this
object were used (Coe \& Orosz, 2000)
and a colour index determined by simply subtracting the red and blue
bands. Figure~\ref{fig:col} shows the result of folding the colour
data at the period of 0.71d and compares the result to the blue band
data folded in the same manner. Also shown in the top panel of this
figure is the best fit sine wave to the colour data. This sine wave
was fitted with no constraints on its amplitude or phase. The
amplitude of this sine wave is 0.0006 magnitudes and the phase is such
that the star is slightly bluer by this amount when at its brightest.

\section{Discussion}

\subsection{Binary motion}

The upper limit results presented for the amplitude of any radial
velocity behaviour in SXP 323 (30 - 50km/s) may be compared to the
expected RV values for the 1.4d orbit suggested by Coe \& Orosz
(2000). Assuming no eccentricity, the expected velocities for a
Keplerian orbit of a 15 $M_o$ star and a 1.4 $M_o$ companion are of
the order 480km/s for such an orbit. Implicit in this calculation is
the assumption that the system is not being viewed pole-on. To check
this assumption, the stellar {\it vsini} was determined using the He 4026\AA
~line and the conversion formula presented in Equation 4 from Steele,
Negueruela \& Clark (1999).  This method gives {\it vsini} = 371$\pm$5
km/s for SXP 323, a higher value then any those authors obtained for a
long list of Be stars in their work. It is therefore exceedingly
unlikely that this star is being viewed pole-on. Furthermore the
clearly split Balmer lines indicate that the circumstellar disk is
also far from being viewed pole-on. It is therefore possible to rule
out binary motion with a period of 0.71d as the explanation for the
sinusoidal modulation of the photometric light curve.

Laycock et al (2005) suggest that, based on X-ray activity cycles, the
correct binary period for this system is 108$\pm$18d. This period is
much closer to the expected binary period of $\sim$180d based upon a
rigorous interpretation of the Corbet diagram (Corbet, 1986). Of
course, the actual Corbet diagram exhibits quite a spread of values
and it is quite feasible that the correct binary period is the X-ray
one. Such a period, assuming a circular orbit, implies speeds of
$\sim$120km/s, which again exceeds the observational upper limits
presented here. A
period of $\sim$180d and a circular orbit 
would give radial velocity amplitudes of $\sim$100km/s.

However, most Be/X-ray binaries present very eccentric orbits and the
presence of regular outbursts in a wide orbit (such as a 108-d or
180-d period would imply) is generally associated with high
eccentricity (cf.  Okazaki \& Negueruela 2001). In such a situation, we
should have observed the source close to periastron in order to see
significant velocity changes, but the data presented here cover only
51 d, which is less than half the X-ray period. Therefore the absence
of obvious radial velocity variations within the 30 km/s limit
reported here does not rule out such a wide orbit. It makes, though,
exceedingly unlikely that the true binary period could be as short as
0.71 d.

\subsection{V/R variability}

From the measured V/R variability presented in Figure~\ref{fig:rvhb}
it is possible to estimate the time for a complete cycle of V/R
changes to be at least $\sim$ 0.3yr. Such timescale does not
difer significantly from those seen in other Be/X-ray binaries (e.g.,
Negueruela et al. 1998). SXP 323 is thought to be a B1
star (Coe et al 2002) and, though there is not a strong variation in
V/R timescales with spectral class (Mennickent \& Vogt, 1991), such a
value would be consistent with other measurements of stars of a similar
type. The presence of obvious V/R variability in the profiles of emission
lines also argues strongly against a close binary. The existence of
V/R variability in Be stars is explained by the theory of global one
armed oscillations in circumstellar disks (Okazaki 2000). 

The presence of global one-armed oscillations requires the existence
of a large disk around the Be star, extending to several $R_{*}$
(Okazaki 2000). Such disk would not have space to 
form in a close binary with a period of 0.7 - 1.4d.

\subsection{NRP}

Previous authors have shown that Be stars exhibiting cyclic
photometric variations show a weak colour variation with phase. Spear,
Mills \& Snedden (1981) studied the Be star 28 Cygni (spectral type
B3Ve) and determined a (U-B) variation of $\sim$0.01 mags in phase
with the 0.7d cycle. As in the observations presented here, the star
got slightly bluer when it got brighter, but the magnitude of the
change in 28 Cyg seems somewhat larger than that seen in SXP 323
(red-blue $\sim$0.001 mags.  However, Spear, Mills \& Snedden (1981)
concluded that they were not seeing NRP behaviour but a phenomenom
related to the rotation of the photosphere.

The case for NRP behaviour is, unfortunately, not supported by evidence
of line profile changes in these VLT data. However, this effect may well be
masked by the changes in the circumstellar disk as indicated by the
evolution of the V/R ratio in H$\beta$ (Figure~\ref{fig:rvhb}). A data
set taken over a much shorter period of time ($\sim$few days) could be
much more successful in exploring this possibility.

\subsection{Interstellar lines}

The multiple structure seen in the interstellar lines is indicative of
the complex spatial structure of the SMC. 

Wayte (1990) obtained CaII absorption spectra of 17 stars in the SMC and
presented their profiles. It is clear from studying their results that
there is a huge range of variations in these profiles across the SMC
reflecting its complex structure. The nearest object to SXP 323 in
the data of Wayte is Sk 35 which reveals features at a heliocentric
velocity of $\sim$50, $\sim$115, $\sim$160 and $\sim$190km/s. Though
some of these are close to our observed features, there are
significant differences as well (for example, our data show nothing at
115km/s). This, though, may not be too surprising given the large
spread of velocity features seen by Wayte across the SMC.

Danforth et al (2002) used FUSE data on 37 objects to identify two
major features: one at 125km/s and another at 156km/s. Both of those
features are probably present in our data, but at slightly different
velocites (132km/s and $\sim$167km/s). The same authors also report a
feature at $\sim$180km/s often seen in the Bar - where SXP 323
lies. This velocity lies between two of the main components reported
here at 167km/s and 201km/s.

Most authors agree that there are major separate structures throughout
the SMC and our data reflect this level of detail. Studies of many of
the other 40-50 SXP sources will add vital information to this complex
question.

\section{Conclusions}

The VLT observations reported here conclusively rule out the
possibility of any binary motion as an explanation for the 0.71d
period seen in the optical counterpart to SXP 323. This fact combined
with the slight colour change seen in the photometric data suggests
that instead this is a phenomenum associated with the Be star - either
NRP or the rotational modulation of photospheric structures. 
However, the probable stability, and
lack of any phase change in the entire MACHO and OGLE data sets of
many years (Coe et al 2002) strongly supports the the NRP case. Future
observations over a much shorter time base should be able to confirm
this directly.

In addition, the spectra show clear evidence for major changes in the
(V/R) ratio of the double peaks of the Balmer lines indicative of
asymmetries in the circumstellar disk. Finally, the complex structure
of the interstellar lines present clear evidence for multiple
structures within the SMC.

\section{Acknowledgements}

Most of the results presented here are based on observations collected
in Service Observing mode at the European Southern Observatory, Chile
under programme 71.D-0093(A).  IN is a researcher of the programme
{\em Ram\'on y Cajal}, funded by the Spanish Ministerio de Educaci\'on
y Ciencia and the University of Alicante, with partial support from
the Generalitat Valenciana and the European Regional Development Fund
(ERDF/FEDER).  This research is partially supported by the Spanish MEC
under grant AYA2002-00814.

\bsp

\label{lastpage}

\end{document}